\documentclass[journal]{IEEEtran}
\usepackage[ruled]{algorithm2e}
\usepackage{graphicx}
\usepackage{epstopdf}
\usepackage{subfig}
\usepackage{amsmath}
\usepackage{amssymb}
\usepackage{color}
\usepackage{multirow}
\usepackage{array}
\usepackage{threeparttable}
\usepackage{caption,booktabs}
\usepackage{hyperref}
\usepackage{enumitem}
\usepackage{algorithmic}
\usepackage{authblk}

\ifCLASSOPTIONcompsoc
  \usepackage[nocompress]{cite}
\else
  \usepackage{cite}
\fi

\begin{document}
\title{Adaptive Similarity Function with Structural Features of Network Embedding for Missing Link Prediction}

\author{
 Chuanting~Zhang$^{1}$, Ke-ke Shang$^{2}$, and Jingping Qiao$^{3}$ \\

    \IEEEauthorblockA{{1} Computer, Electrical and Mathematical Science Engineering Division, King Abdullah University of Science and Technology (KAUST), Thuwal 23955, Saudi Arabia\\
    }
    \IEEEauthorblockA{{2} Computational Communication Collaboratory, Nanjing University, Nanjing 210093, China\\
    }
    \IEEEauthorblockA{{3} School of Information Science and Engineering, Shandong Normal University, Jinan 250022, China\\
    }

	\thanks{Corresponding authors: Ke-ke Shang (kekeshang@nju.edu.cn) and Jingping Qiao ( jingpingqiao@sdnu.edu.cn). }
}

\maketitle

\begin{abstract}
Link prediction is a fundamental problem of data science, which usually calls for unfolding the 
mechanisms that govern the micro-dynamics of networks.
In this regard, using features obtained from network embedding for predicting links has drawn widespread attention.
Though edge features based or node similarity based methods have been proposed to solve the link prediction problem, many technical challenges still exist due to the unique structural properties of networks, especially when the networks are sparse.
From the graph mining perspective, we first give empirical evidence of the inconsistency between heuristic and learned edge features.
Then we propose a novel link prediction framework, \textit{AdaSim}, by introducing an Adaptive Similarity function using features obtained from network embedding based on random walks.
The node feature representations are obtained by optimizing a graph-based objective function.
Instead of generating edge features using binary operators, we perform link prediction solely leveraging the node features of the network.
We define a flexible similarity function with one tunable parameter, which serves as a penalty of the original similarity measure.
The optimal value is learned through supervised learning thus is adaptive to data distribution.  
To evaluate the performance of our proposed algorithm, we conduct extensive experiments on eleven disparate networks of the real world.
Experimental results show that \textit{AdaSim} achieves better performance than state-of-the-art algorithms and is robust to different sparsities of the networks.
\end{abstract}

%\begin{IEEEkeywords}
%Link Prediction, Adaptive Similarity Function, Network Embedding, Representation Learning, Network Analysis.
%\end{IEEEkeywords}

% \fi
\IEEEpeerreviewmaketitle

\section{Introduction}\label{sec:introduction}
%说明复杂网络研究的重要性
\IEEEPARstart{N}{etworks} have recently emerged as an important tool for representing and analyzing many kinds of interacting systems ranging from biological to social science\cite{Newman:2010:NI:1809753}.
As technological innovation and data explosion gather pace, we humans are now moving into the era of big data, hence the reach and participate of these networks is rapidly expanding.
Studying these complex, interlocking networks can help us understand the operation mechanism of real-world systems.
Therefore, in the past years, lots of work has been dedicated to studying evolution\cite{Palla2007,Leskovec2007}, topologies\cite{Newman2003,Newman2016}, and characteristics\cite{Lue2015} of networks, attracting researchers from physics, sociology, and computer science.

%引出网络在很多情况下是不完整的，存在信息（边）的缺失，需要进行预测，并有很重要的应用
Under many circumstances however, the current observations of various network data are substantially incomplete\cite{Clauset2008}.
For example, in protein-protein interaction and metabolic networks, whether two nodes have a link must be determined experimentally, which is very costly.
As a result, the known links may represent fewer than 1\% of the actual links\cite{Soundarajan2012}.
Besides, in social networks like Facebook, only part of the friendships among users are shown by the observed network, there still exist user pairs who already know each other but are not connected through Facebook.
Due to this, it is always a challenging yet meaningful task to identify which pairs of nodes not connected in the current network are likely to be connected in the actual network, i.e., predicting missing links.
Acquiring such knowledge is useful, for example, in biological domain, it gives invaluable guidance to carry out targeted experiments, and in social network domain, it can be used to recommend promising friendships, thus enhance users' loyalties to web services.

%解决链接预测的方法分类主要有监督学习方法和非监督学习方法，引出监督方法取得了更好的结果
The way to solve the link prediction problem\cite{Shang2017,Shang2019,Wu2020,Wang2020a,Rossi2021,Cai2021} can be roughly divided into two categories, i.e., unsupervised methods and supervised methods.
In current research work on unsupervised link prediction, they mainly focus on defining a similarity metric $s_{uv}$ for unconnected node pairs $(u,v)$ using information extracted from the network topology.
The defined metrics represent different kinds of proximity between a pair of nodes and have different performance among various networks and no one can dominate others.
Most of the metrics are easy to compute and interpret, but they are so invariant that fundamentally unable to cope with dynamics, interdependencies, and other properties in networks\cite{Lichtenwalter2010}.
Machine learning and artificial intelligence technologies \cite{zhang2018citywide,zhang2019deep,Shen2021,Zhang2021a,Zhang2021} are revolutionizing many domains including graph mining. Thus, the link prediction problem can also be posed as a supervised binary classification task from a machine learning perspective\cite{al2006}.
Since then the research of supervised methods for link prediction has become prominent  \cite{Lichtenwalter2010, Backstrom:2011, Davis2013, Ade2016}, and the results of these researches provide confirmatory evidence that a supervised approach can enhance the link prediction performance.

%监督学习中很重要的一个地方就是特征选取，引出利用自动学习的特征做链接预测
Choosing an appropriate feature set is crucial for any supervised machine learning task \cite{Chang2016,Wang2020,Yuan2021}.
For link prediction, each sample in the dataset corresponds to a pair of nodes.
A typical solution is using multiple topological similarities as features and this is the most intuitive way.
But all these features are handcraft and cost much human labor.
Besides, they often rely on domain knowledge, thus restrict the generalization across different fields.

An alternative method is learning the features automatically for the network.
By treating networks as special documents consist of a series of node sequences, the node features can be learned by solving an optimization problem\cite{Perozzi-2014}.
After obtaining the features of nodes, the link prediction task is traditionally conducted using two approaches.
The first one is similarity-based ranking method\cite{Zhiyuli2016}, for example, cosine similarity is used to measure the similarity of pairs of nodes.
For two unconnected nodes, the larger the similarity value, the higher the connection probability they have.
The other one is edge feature based classification method\cite{Zhang2016,Grover2016}.
In this method, the edge features are generated by heuristic binary operators such as \textit{Hadamard} operator and \textit{Average} operator.
Then a classifier is trained using these features and will be used to distinguish whether a link will form between two unconnected nodes.

\begin{figure*}[!ht]
\centering
\subfloat[][]{\includegraphics[width=0.33\textwidth]{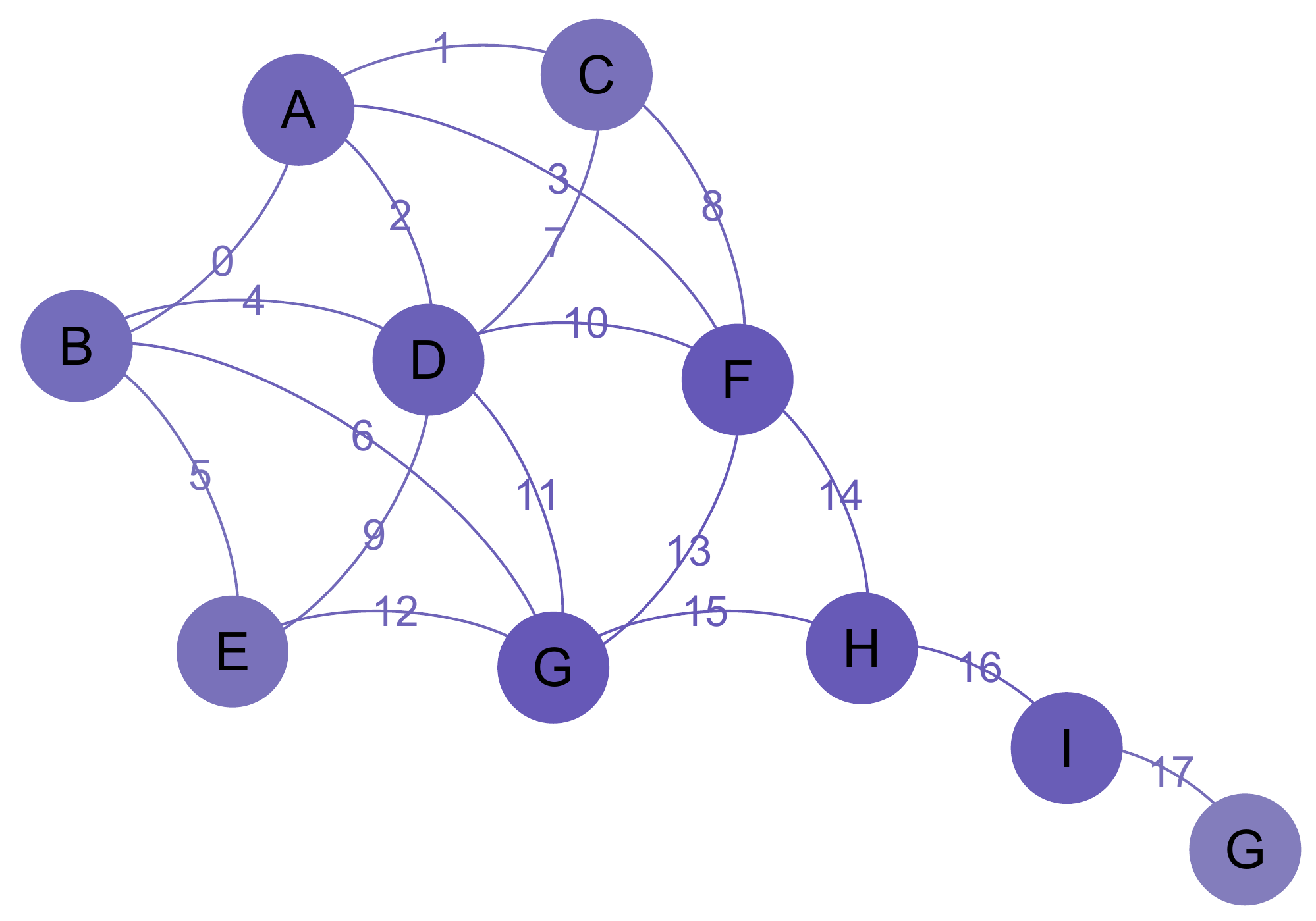}\label{toy_graph}}
\subfloat[][]{\includegraphics[width=0.66\textwidth]{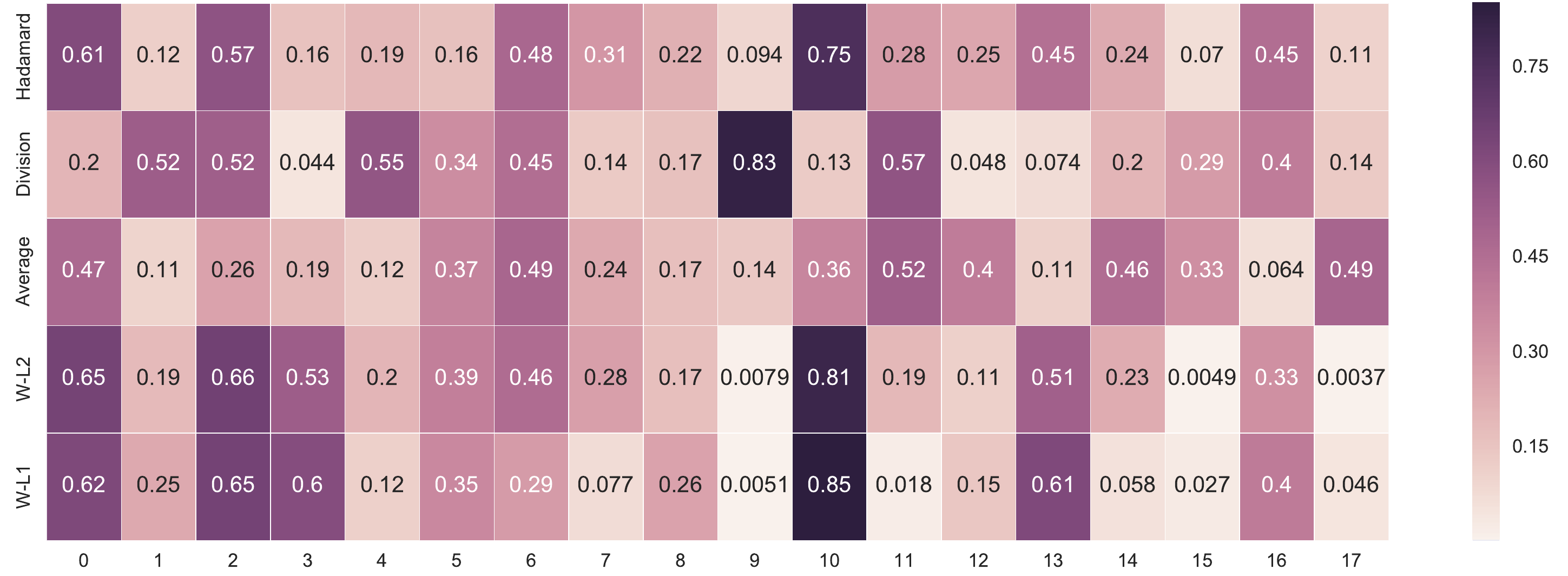}\label{edge_coeff}}
\caption{(a)A toy network.~(b)Pearson correlation coefficients between learned edge features and heuristic edge features. The node sequences sampling strategy used in this example is random walk.}
\label{graph_and_coeff}
\end{figure*}

%指出现在两种方法/框架的局限性和不足
As the features learned through network embedding preserve the network's local structure, the cosine similarity works well for strongly assortative networks but fails to capture the disassortativity of the network, i.e., nodes prefer to build connections on large scales than on small scales\cite{Clauset2008}.
Thus using cosine similarity for link prediction suffers from statistical performance drawbacks.
Besides, the edge features obtained through binary operators will potentially lose node's information, since the features of nodes are learned by solving an optimization problem but the edge features are not (See Fig.\ref{graph_and_coeff} for a clear explanation and the details will be discussed in Section \ref{evidence}).
Furthermore, the edge and node features have the same dimensionality, which is usually on the scale of several hundred.
This means that even for linear models such as logistic regression, it still needs to learn hundreds of parameters, which presents us with the question of feasibility especially when the data size is large.
How to design a simple, general yet efficient link prediction method using the node features directly learned from network embedding still remains an open problem.

%本文的主要创新点介绍
To solve the above mentioned issues, we propose a novel link prediction method, \textit{AdaSim} (\underline{Ada}ptive \underline{Sim}ilarity function), for large scale networks.
The node feature representations are obtained by optimizing a graph-based objective function using stochastic gradient descent techniques.
Instead of generating edge features using heuristic binary operators, we perform link prediction solely leveraging the node features of the network.
Our essential contribution lies in defining a flexible node similarity function with only one tunable parameter, which serves as a penalty of the original similarity.
The optimal value can be obtained through supervised learning thus is adaptive to the data distribution, which gives \textit{AdaSim} the ability to capture the various link formation mechanisms of different networks.
Compared with the original cosine similarity, the proposed method generalizes well across various network datasets.

In summary, our main contributions are listed as follows.
\begin{itemize}

\item We propose, \textit{AdaSim}, a novel link prediction method by introducing an adaptive similarity function using features learned from network embedding.
\item We show that \textit{AdaSim} is flexible enough with only one tunable parameter. It is adjustable with respect to the network property. This flexibility endows \textit{AdaSim} with the power of capturing the link formation mechanisms of different networks.

\item We demonstrate the effectiveness of \textit{AdaSim} by conducting experiments on various disparate networks of the real-world. The results show that the proposed method can boost the performance of link prediction in different degrees. Besides, we find that \textit{AdaSim} works particularly well for highly sparse networks.
\end{itemize}

The rest of the paper is structured as follows. 
Section \ref{related-work} reviews some research works related to link prediction.
The problem definition of link prediction and feature learning are described in section \ref{problem-statement}, and some empirical findings on the datasets are also given in this section. 
Section \ref{proposed-framework} illustrates the proposed link prediction method \textit{AdaSim} with detail explanations of each component. 
The experimental results and analysis are represented in Section \ref{experiments}. 
Finally, Section \ref{conclusion} concludes the paper.

\section{Related work}\label{related-work}
%相关工作介绍。首先介绍非监督方法，然后监督学习方法（同质网络和异质网络），进而着重介绍如何用自动学习得到的特征做链接预测。最后一段指出本文工作与之前工作的不同。
Early works on link prediction mainly focus on exploring topological information derived from graphs.
Liben-Nowell and Kleinberg\cite{Liben-Nowell:2007} studied several topological features such as common neighbors, Adamic-Adar, PageRank and Katz and found that topological information is beneficial in predicting links compared with a random predictor.
Subsequently, some topology-based predictors were proposed for link prediction, e.g., resource allocation\cite{Zhou2009}, community-enhanced predictors\cite{Soundarajan2012} and clustering coefficient-based link prediction\cite{Wu20161}.

Hasan \textit{et al.} were the first to model link prediction as a binary classification problem from a machine learning perspective\cite{al2006}.
Various similarity metrics between node pairs are extracted from the network and treated as features in a supervised learning setup, then a classifier is built with these features as inputs to distinguish positive samples (links that form) and negative samples (links that do not form).
Thereafter the supervised classification approach has been prevalent in the link prediction domain.
Lichtenwalter \textit{et al.} proposed a high-performance link prediction framework called HPLP. Some new perspectives for link prediction, e.g., the generality of an algorithm, topological causes and sampling approaches, were included in \cite{Lichtenwalter2010}.
Later, a supervised random walk-based algorithm was proposed by Backstrom and Leskovec\cite{Backstrom:2011} to effectively incorporate the information from the network structure with rich node and edge attribute data.

In addition, the link prediction problem is also extended to heterogeneous information networks\cite{sun5992571, Dong6413904, Dong:2015:CLP, zhang2015integrated}.
Among these works, a core concept based on network schema was proposed, namely meta-path.
Multiple information sources can be effectively fused into a single path and different meta paths have different physical meanings.
Some similarity measures can be calculated using meta paths, then they are treated as features of a classifier to discriminate positive and negative links.

All the works mentioned above on supervised link prediction use handcraft features, which require expensive human labor and often rely on domain knowledge.
To alleviate this, one can use the latent features learned automatically through representation learning\cite{Bengio2013}.
For networks, the unsupervised feature learning methods typically use the spectral properties of various matrix representations of graphs, such as adjacency and  Laplacian matrices.
In the perspective of linear algebra, this kind of method can actually be regarded as a dimensional reduction technique.
Several works\cite{Roweis2323, belkin2003laplacian} have been done aiming to acquire the node features of graphs, but the computation of eigendecomposition of a matrix is costly, thus makes these methods impractical to scale up to large networks.

Perozzi \textit{et al.}\cite{Perozzi-2014} extended the skip-gram model to graphs and proposed a framework, DeepWalk, by representing a network as a special ``document'' consists of a series of node sequences, which are generated by random walks.
DeepWalk can learn features for nodes in the network, and the representation learning process is irrelevant to downstream tasks like node classification and link prediction.
Later, Node2vec was proposed by Grover and Leskovec\cite{Grover2016}.
Compared with DeepWalk, Node2vec uses a biased random walk to control the sampling space of node sequences.
The network properties such as homophily and structure equivalence can be captured by Node2vec.
The link prediction was performed using edge features obtained through heuristic binary operators on node features.
In \cite{Wang:2016:SDN}, the authors proposed a deep model called SDNE to capture the highly non-linear property of networks.
The first-order and second-order proximity were jointly exploited to capture the local and global network structure, respectively.
More recently, Wang \textit{et al.}\cite{wangaaai2017} proposed a novel Modularized Nonnegative Matrix Factorization (M-NMF) model to incorporate not only the local and global network structure but also the community information into network embedding.
In order to model the diverse interacting roles of nodes when interacting with other nodes, Tu \textit{et al.}\cite{Tu2017} presented a Context-Aware Network Embedding (CANE) method by introducing a mutual attention mechanism.
CANE can model the semantic relationships between nodes more precisely.
In order to save computation time, in\cite{Zhiyuli2016}, link prediction was directly carried out using cosine similarity of node features instead of edge features.

The above works mainly focus on the network embedding techniques and ignore the typical characteristics of link formation.
The main difference between existing work and our efforts lies in that we consider an adaptive similarity function yet with a learning-based idea, making our model flexible enough to capture the various link formation patterns of different networks. For example, a negative value of $p$ can weaken the role of `structural equivalence' and enhance the score of dissimilar node pairs, thus capturing the disassortativity on link formation.

%***********************************Figure BEGIN***************************************
%***********************************Figure END*****************************************

\section{Problem statement and feature learning framework}\label{problem-statement}
In this section, we first give the formal definition of the link prediction problem.
Then the feature learning framework for networks is presented.
Finally, we introduce the empirical findings on several network datasets when using node features for link prediction.

\subsection{Problem Formulation}
\label{subsec:pf}
Given a network $G=(V,E)$, where $V$ is the set of nodes, $E \in (V \times V)$ is the set of links.
No multiple links or self-links are allowed for any two nodes in the network.
It is assumed that some of the links in the network are unobserved or missing at the present stage.
The link prediction task aims to predict the likelihood of a link between two unconnected nodes using information intrinsic to the network.

Since here we are considering a supervised approach for link prediction, we first need to construct a labeled dataset $D=\{(\mathbf{x}_i, y_i), i \in [1,m]\}$, where $\mathbf{x}_i$ is the feature vector of the $i\text{-th}$ sample and $y_i \in \{0,1\}$ the corresponding label.
Specifically, $\mathbf{x}_i = \Phi (\mathbf{u}_i, \mathbf{v}_i)$ in which $\mathbf{u}_i$ and $\mathbf{v}_i$ denote the features of node $u_i$ and $v_i$, respectively.
The node features are learned from network representation learning.
$\Phi (\cdot)$ is a mapping function from node features to node pair features.
For any node pair in $D$, $y_i = 1$ indicates that this node pair belongs to positive samples and otherwise the negative samples.
Positive samples are the edges, $E^p$, chosen randomly from the network $G$.
We delete $E^p$ from G and keep the obtained sub-network ($G_s$) is fully connected.
To generate negative samples, we sample an equal number of node pairs from $G$ which having no edge connecting them.
The dataset $D$ is spitted into two parts: training dataset $D_T$ and test dataset $D_P$.
A classification model $\mathcal{M}$ can be learned with dataset $D_T$, then this model will be used for predicting whether a pair of nodes in dataset $D_P$ should have a link connecting them. Our algorithms are typical methods from the field of graph mining. Hence, in contrast to part of our previous papers\cite{Shang2017,Shang2019}, we follow conventions in the field of artificial intelligence in which $E^p$ (positive samples) has 50\% of the observed links, and scores are based on the combination of $E^p$ and the same number of non-observed links (negative samples). For the highly sparse sexual contact network, which has only a small number of nodes, $E^p$ instead comprises all observed links.

\subsection{Feature learning of network embedding}\label{language_model}  
For a given network $G=(V, E)$, a mapping function $f:V\longrightarrow \mathbb{R}^{|V| \times d}$ from nodes to feature vectors can be learned for link prediction.
Here $d$ is a user-specified parameter that denotes the number of dimensions of the feature vectors and $f$ is a matrix of size $|V| \times d$ parameters.
The mapping function $f$ is learned through a series of document-like node sequences, using optimization techniques originated in language modeling.

The purpose of language modeling is to evaluate the likelihood of a sentence appearing in a document.
The model is built using a corpus $\mathcal{C}$.
More formally, it aims to maximize
\begin{equation}
Pr(w | context(w))
\end{equation}
over all training corpus, where $w$ is a word of the vocabulary, $context(w)$ is the context of $w$ that includes the words that appear to both the left side of $w$ and the right side.
Recent research on representation learning has put a lot of attention on leveraging probabilistic neural networks to build a general representation of words, extending the scope of language modeling beyond its original goals.
Each word is represented by a continuous and low-dimensional feature vector.
The problem then, is to maximize
\begin{equation}
Pr(w | (f(w_{i-j}),\cdots,f(w_{i-1}),f(w_{i+1}),\cdots,f(w_{i+j}))),
\end{equation}
where $f(\cdot)$ denotes the latent representation of a word.

The social representation of networks can be learned analogously through a series of node sequences generated by a specific sampling strategy $\mathcal{S}$.
Similar to the context of word $w$ in language modeling, $N_{\mathcal{S}}(u)$ is defined to be the neighborhood of node $u$ using sampling strategy $\mathcal{S}$.
The node representation of networks can be obtained by optimizing the following expression
\begin{equation}
Pr(u | (f(u_{i-j}),\cdots,f(u_{i-1}),f(u_{i+1}),\cdots,f(u_{i+j}))).
\end{equation}
The learned representations can capture the shared similarities in local graph structure among nodes in the networks.
Nodes that have similar neighborhoods will acquire similar representations.

\subsection{Empirical findings on several network datasets}\label{evidence}
%这一部分主要是对用“二元操作子得到边特征”和直接“用相似度”做链接预测的不足给出证据。
After learning the representations for the nodes in the network, there are two approaches to the link prediction task, i.e., node similarity-based method and edge feature-based method.
The former is simple and scalable, and the latter is complex yet powerful.
But both methods have their limitations in effectively characterizing the link formation patterns of node pairs.
Since the node similarity-based method was not involved in learning, it can not be aware of the effects of global network property in link prediction.
The edge feature-based method could not describe the node pair relationship very well at the feature level using a heuristic binary operator, as the information loss exists in the mapping procedure from node features to edge features.
We show the empirical evidence for the limitations of these two kinds of methods in the following subsections.

\begin{table}[!ht]
\renewcommand{\arraystretch}{1.5}
\centering
\caption{Choice of binary operators for learning edge features.} 
%The definitions correspond to the $i$th instance in dataset $D$.}
\label{binary_operators}
\begin{tabular}{c|c|l}
\toprule
\textbf{Operator} & \textbf{Symbol}   & \multicolumn{1}{c}{\textbf{Definition}}   \\
\midrule
Hadamard & $\odot$ & $[f(u)\odot f(v)]_i=f_i(u)*f_i(v)$ \\
Average & $\oplus$ & $[f(u)\oplus f(v)]_i=\frac{f_i(u)+f_i(v)}{2}$ \\
Division & $\oslash$ & $[f(u)\oslash f(v)]_i=\frac{f_i(u)}{f_i(v)}$ \\                                       
Weighted-L1 & $\Vert \cdot \Vert_{\bar{1}}$ & $[f(u)\cdot f(v)]_i=|f_i(u)-f_i(v)|$ \\
Weighted-L2 & $\Vert \cdot \Vert_{\bar{2}}$ & $[f(u)\cdot f(v)]_i=|f_i(u)*f_i(v)|^2$ \\
\bottomrule
\end{tabular}
\end{table}

%***********************************Figure BEGIN***************************************
\begin{figure*}[!ht]
     \centering
     \subfloat[][Epinions]{\includegraphics[width=0.33\textwidth]{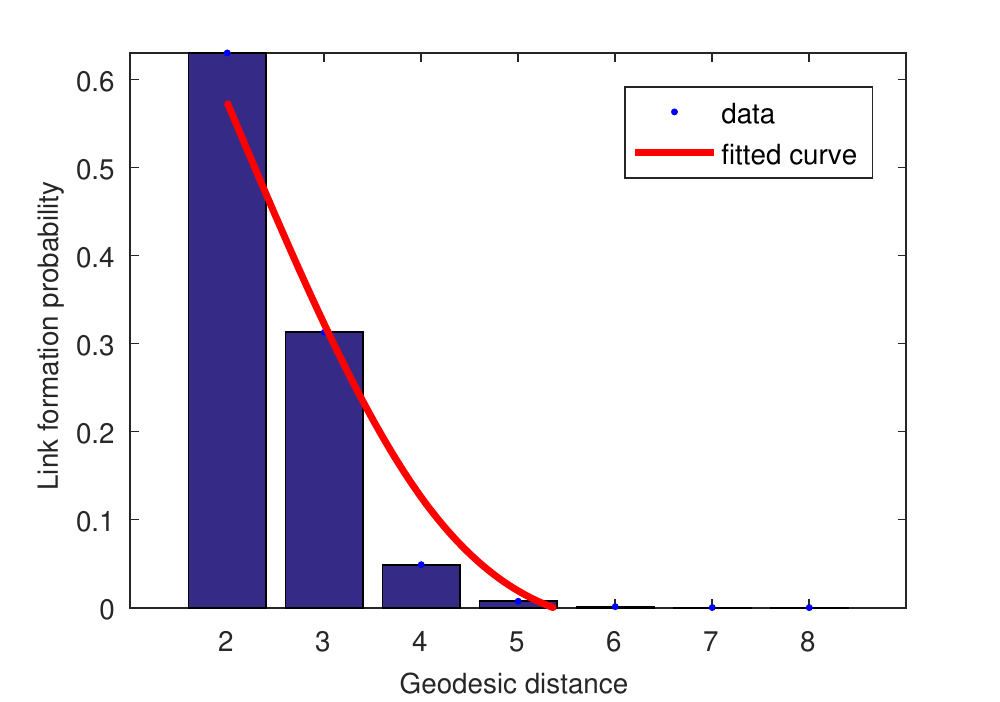}\label{pattern1}}
     \subfloat[][Gnutella]{\includegraphics[width=0.33\textwidth]{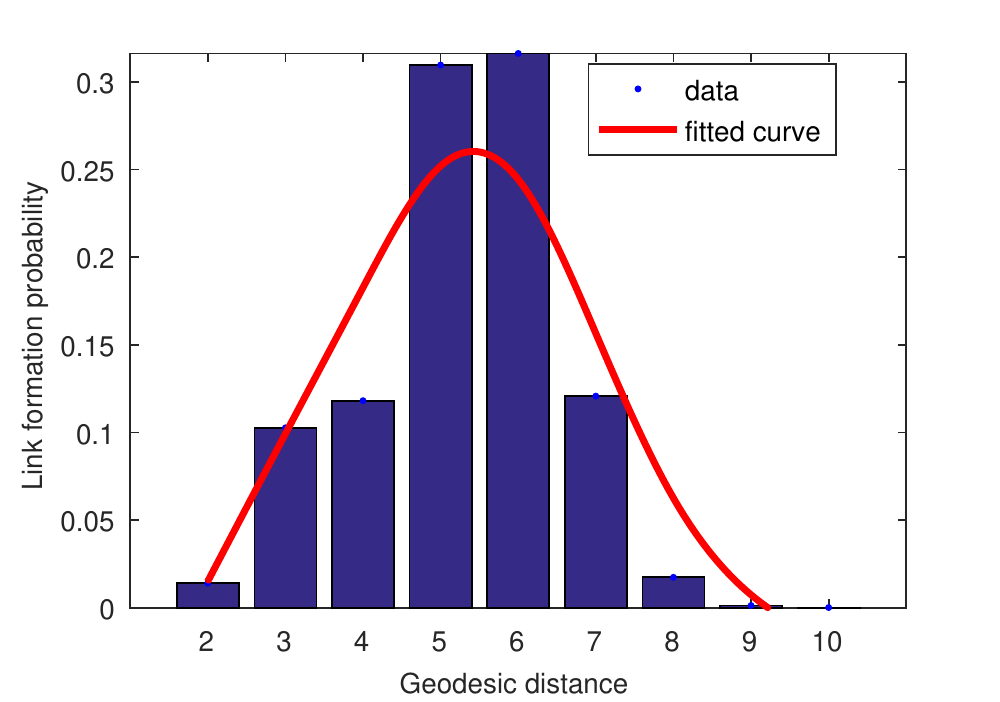}\label{pattern2}}
     \subfloat[][Router]{\includegraphics[width=0.33\textwidth]{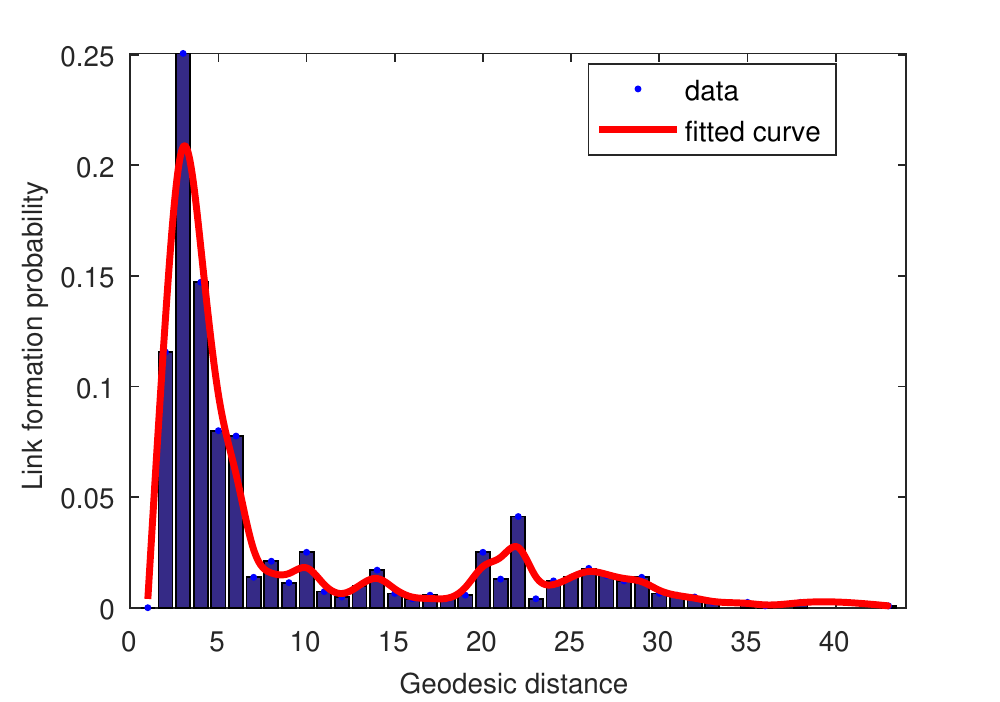}\label{pattern3}}
     \caption{Link formation patterns among different networks. The $x$ axis represents the geodesic distance $s$ between a pair of nodes $(u, v)$ and the $y$ axis represents the link formation probability. The probabilities are calculated by $|E^p_s|/|E^p|$ in which $|E^p|$ is the number of positive links that are randomly sampled from the network, and $|E^p_s|$ is the number of pairs of nodes that their geodesic distance is exactly $s$.}
     \label{link_mechanism}
\end{figure*}
%***********************************Figure END****************************************
\subsubsection{Limitation of heuristic binary operators}
%传统的研究中，通过特征学习得到节点的特征，然后利用启发式二元操作子得到边的特征。然而，这样得到的边特征会丢失部分节点的信息，我们给出了实际数据的证据。首先，以特征学习的方式得到边的真实特征表达，然后以二元操作子得到这边边的“启发式”表达。如果启发式的边特征能够很好的保留节点信息，那么启发式边特征与真实的边特征应该有很高的皮尔逊相关系数。但是，实际的结果并不是这样的，图1所示。
Fig.\ref{toy_graph} shows a toy network (krackhardt kite graph) with 10 nodes and 18 edges.
Each node and edge is marked with a unique label.
Given a specific sampling strategy $\mathcal{S}$, we can obtain the node sequences and the corresponding edge sequences simultaneously after performing $\mathcal{S}$ on the network.
Hence, both node representations and edge representations of the network can be learned using optimization techniques.
For a specific pair of nodes, the learned edge representation is called the ``true'' features and the generated edge representation using binary operator is call the ``heuristic'' features.
It is known that if a binary operator is good enough, it should be able to accurately characterize the relationship of pairs of nodes, i.e., the correlation between heuristic features and true features should be as strong as possible. 
For the 18 edges in the toy network, five different kinds of heuristic binary operators\cite{Grover2016} (see Table \ref{binary_operators}) are chosen to generate edge features\footnote{For the \textit{Division} operator, we omit the kind of $\frac{f_i(v)}{f_i(u)}$ since it has very similar results compared with $\frac{f_i(u)}{f_i(v)}$.}, and their correlation with the true edge features are displayed in Fig.\ref{edge_coeff}.

On the basis of the evidence from Fig.\ref{graph_and_coeff}, we can tell that different operators have different results in representing features of pairs of nodes and no one can dominate the others.
Some of the edges, e.g., edge 10 and 16, can be well characterized by the \textit{Hadamard} operator, while others, for example, edge 12, 14 and 17, can be characterized by the \textit{Average} operator.
Furthermore, most values are less than 0.5, which means a weak correlation between the heuristic edge features and true edge features.
This verifies our claim that edge features obtained through heuristic binary operators may cause the loss of information of node features.
%some of the information of node features are not accurately captured by the binary operator.
\subsubsection{Limitation of similarity based method}
Given an unconnected node pair $(u,v)$, several metrics can be used to measure their similarity, for example, the common neighbors between $u$ and $v$, the number of reachable paths from $u$ and $v$.
But here we only consider the metric of cosine similarity since we have the node pair's feature vectors $\mathbf{u}$ and $\mathbf{v}$, respectively.
The cosine similarity is used to characterize the link formation probability and it is defined as
\begin{equation}
cos(u,v)=\frac{\mathbf{u}^T \mathbf{v}}{\Vert \mathbf{u}\Vert \Vert \mathbf{v}\Vert},
\end{equation}
where $(\cdot)^T$ denotes the transpose and $\Vert \cdot \Vert$ means the $l_2$-norm of a vector. 
The cosine similarity measures the cosine of the angle between two $d$-dimensional vectors obtained from network representation learning.
In fact the idea of cosine similarity has been used for link prediction in several works\cite{Lu2011,gao2015,Zhiyuli2016}.
But there are a few issues when directly using cosine similarity for link prediction.
The first one is it did not consider the label information of node pairs. Thus it belongs to the category of unsupervised learning.
However, lots of works have demonstrated that supervised learning approaches to link prediction can enhance the performance\cite{Lichtenwalter2010, Davis2013, Ade2016}.
The other one is cosine similarity is too rigid to capture different link formation mechanisms of different networks.

%在特征学习的阶段，有一个假定就是：在网络中，两个离的越近的节点，节点之间的余弦相似性就越大。然而，在异配网络中，这个假设并不成立。因为，异配网络中，节点更倾向于与离自己远的节点产生链接。我们通过实际的数据观察给出证据（empirical evidence）。
Since in the phase of representation learning for networks, it is assumed that two nodes have similar representations if they have similar context in the node sequences sampled by strategy $\mathcal{S}$.
For networks, this indicates that if two nodes are structurally close\footnote{For the three nodes $(v_1,v_2,v_3)$ in the graph, suppose the geodesic distance of $(v_1, v_2)$ is 2 and $(v_1, v_3)$ is 5, we say $v_2$ is closer to $v_1$ than $v_3$.} to each other, then they have a high probability to simultaneously occur in the same sequence which results in a high value in terms of cosine similarity.
But in real-world networks, whether two nodes will form a link is not simply influenced by this kind of structural closeness.
Two nodes far from each other in the network will also have a high chance to build relationships if they are structural equivalence\cite{Grover2016}.
For two nodes, ``the closer the graph distance, the easier for them to build link'' holds not necessarily true, especially when the network is sparse and disassortative.

As shown in Fig.\ref{link_mechanism}, we can see that different networks have different patterns in building new connections\footnote{The datasets are described in Section \ref{data}}.
These patterns are closely related to the network properties, such as clustering coefficient, graph density and assortativity.
To some networks with high assortativity, two unconnected nodes tend to be connected if they are structurally close, while others are not.
More specifically, the link formation probability for two unconnected nodes is vastly decreasing with the increase of geodesic distance in the \textit{C.elegans} dataset, and 97.8\% new links span the geodesic distance less than 3.
But for the \textit{Gnutella} dataset, with an increase of geodesic distance, the link formation probability first increases then decreases, and most of the new links (62.4\%) are generated by node pairs with distance equals to 5 or 6.
For the \textit{Router} dataset, the new links span a wide range of geodesic distances from 2 to 34 and almost half of the new links (48.67\%) span a distance larger than 5.
The distribution of link formation probabilities is more complex than the other two datasets.

The cosine similarity function assigns higher scores to pairs of nodes if they are close to each other and vice versa.
It can capture link formation patterns in the case of Fig.\ref{pattern1}, i.e., the shorter the distance between two unconnected nodes, the higher the probability to be connected.
But cosine similarity fails to capture the patterns in the cases of Fig.\ref{pattern2} and Fig.\ref{pattern3}, especially Fig.\ref{pattern2}, in which the link formation pattern follows a Gaussian distribution.
For the pattern of Fig.\ref{pattern2}, nodes prefer to build connections with those that are relatively farther from them.
When performing a link prediction task in cases like this, pairs of nodes with a relatively longer distance should be more similar than those with a shorter one.

%***********************************Figure BEGIN***************************************
\begin{figure*}[]
\centering
\includegraphics[width=1.0\textwidth]{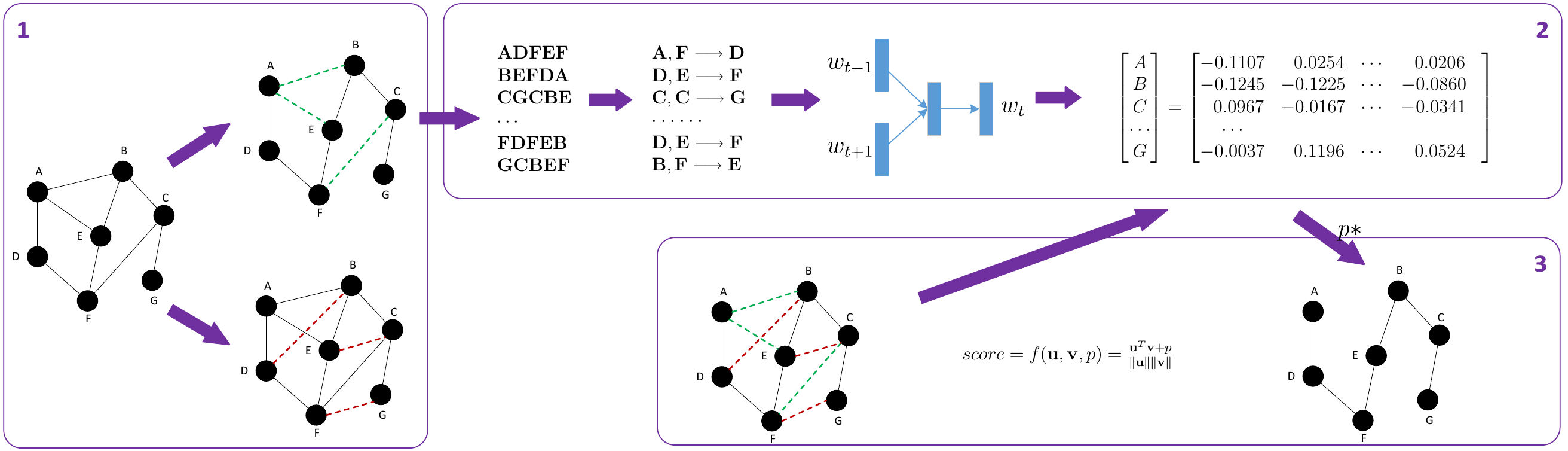}
\caption{The whole framework of link prediction based on adaptive similarity function.}
\label{framework}
\end{figure*}
%***********************************Figure END*****************************************

%因为我们给出了证据，边特征会丢失节点信息，所以，我们直接用节点特征做链接预测。另外，由于传统的基于节点相似度的方法不能很好的应对网络链接模式的多样性（比如：异配网络），所以需要在保证算法简单的情况下，学习出来一个相似度函数，来动态刻画节/调整节点的相似性大小。
Thus, we need to design a flexible similarity function for link prediction and capture the various patterns of link formation.
Besides the similarity function should be devised on the basis of concision and scalability.
This can be achieved by adjusting the similarity of node pairs and balancing link formation probabilities among different distances.
Inspired by this, we propose a modified similarity function which is defined as

\begin{equation}
\text{sim}(u,v)=\frac{\mathbf{u}^T \mathbf{v}}{\Vert \mathbf{u}\Vert \Vert \mathbf{v}\Vert} +
\frac{p}{\Vert \mathbf{u}\Vert \Vert \mathbf{v}\Vert},
\end{equation}
where $p$ is a balance factor to control the similarity of node pairs with different geodesic distances.
As we have the labels of node pairs, so the optimal value of $p$ can be learned in a supervised way.

\section{The proposed framework}\label{proposed-framework}
%整个算法框架：子图提取、子图上的特征学习、参数p的学习
In this work, we propose a novel link prediction framework, \textit{AdaSim}, based on an adaptive similarity function using the features learned from network representation.
The whole framework is illustrated in Fig.\ref{framework}. 
It can be divided into three parts: subgraph generation, feature representation and similarity function learning.
First, the positive and negative node pair indexes are obtained through random sampling.
The corresponding subgraph $G_s$ is generated via edge removal.
Then we learn the representation of nodes in the network using an unsupervised way.
Finally a similarity function is defined and the optimal parameter is determined through supervised learning. 
The obtained similarity function with optimal penalty can be directly used to solve the link prediction problem.
\begin{algorithm}[]
\caption{Sub-graph generation}
\label{alg1}
\begin{algorithmic}[1]
\REQUIRE $G=(V,E)$, positive edge ratio $r$
\ENSURE samples of node pairs and sub-graph $G_s$
\STATE $E_{mst} = \text{Kruskal}(G)$
\STATE $n = |E|*r$
\STATE Shuffle($E - E_{mst}$)
\STATE $E_p = (E - E_{mst})[0:n]$

\FOR{$e\notin E$ and $|E_n| \leqslant n$}
	\STATE append $e$ to $E_n$
\ENDFOR
\STATE $G_s=(V, E-E_p)$
\RETURN $G_s$, $E_p+E_n$
\end{algorithmic}
\end{algorithm}
\subsection{Subgraph generation}
Unlike other tasks such as link clustering or node classification, in which the complete structural information is available, a certain fraction of the links needs to be removed before performing network representation learning for link prediction.
In order to achieve this, one can iteratively select one link and determine whether it is removable or not.
But this operation is less effective and very time consuming especially when the network is very sparse since it needs to traverse almost all the nodes in the graph.

Instead, we propose a fast positive sampling method based on minimum spanning tree (MST) in this paper.
A MST is a subset of the edges in the original graph $G$ that connects all the nodes together.
That means all the edges are removable except those that belong to the MST and their deletion will not break the property of $G$ of being connectivity.
Lines 1--4 in Algorithm \ref{alg1} shows the core of our approach.
We first generate a MST of $G$ denoted as $G_{mst}=(V, E_{mst})$ using Kruskal's algorithm.
The positive samples $E_p$ are randomly selected from $E - E_{mst}$.
To generate negative samples $E_n$, we sample an equal number of node pairs from $G$, with no edge connecting them (lines 5--7).
Then we delete all the edges in $E_p$ from $G$ and obtain the subgraph $G_s$ (line 8).

\begin{algorithm}[]
\caption{Feature representation}
\label{alg2}
\begin{algorithmic}[1]
\REQUIRE $G_s=(V,E-E_p)$, window size $\lambda$, feature size $d$, walks per node $k$, walk length $l$
\ENSURE node representation $f$
\FOR{$i=1$ to $k$}
	\STATE $\text{Shuffle}(V)$
	\FOR{$v \in V$}
		\STATE walks = RandomWalk($G_s$, $v$, $l$)
	\ENDFOR
\ENDFOR
\FOR{walk $\in$ walks}
	\FOR{$v$ $\in$ walk}
		\STATE{$J(f)=-log~Pr(v|\mathbf{x}_v)$}
		\STATE{$f = f - \alpha * \frac{\partial J}{\partial f}$}
	\ENDFOR
\ENDFOR
\RETURN $f$
\end{algorithmic}
\end{algorithm}

\subsection{Feature representation}
Now we proceed to perform the feature learning task on subgraph $G_s$.
This task consists of two core components, i.e., a node sequence sampling strategy and a language model.
\subsubsection{Node sequence sampling}

In terms of node sequence sampling, the most classical strategies are Breadth First Search (BFS) and Depth First Search (DFS)\cite{Grover2016}.
BFS starts at a specific node and explores the neighbors first before moving to the next level.
Contrarily, DFS traversing the network starts at one node and explores as far as possible along each branch before backtracking.
BFS and DFS represent two extreme sampling strategies with respect to the search space they explore, bringing about valuable implications on the learned representations.
In fact, the neighborhood sampled by BFS can reflect the structural equivalence about the networks and the sampled nodes in DFS can reflect a macro-view of the neighborhoods, which is essential in inferring communities based on homophily\cite{Grover2016}. 
Though they are of paramount significance for producing interesting representations, neither can simultaneously reveal the complex properties of networks.
We need a sampling strategy that can smoothly interpolate between DFS and BFS, whose requirement can be fulfilled by random walks on graphs.

A random walk of length $l$ on $G_s$ rooted at node $u$ is a stochastic process with random variables $(v_1,v_2,\cdots,v_k)$ such that $v_1=u$ and $v_{i+1}$ is a node chosen uniformly at random from the neighbors of $v_i$.
Random walks arise in a variety of models for large scale networks, such as computing node similarities\cite{Fouss2007, Ade2016}, learning to rank nodes\cite{Yu2014,Chen2015} and estimating network properties\cite{Cooper2013}.
Besides, they are the foundation of a class of output-sensitive algorithms that employ them to calculate community structure's local information.

This connection is the reason that motivates us to use random walks as the node sequence sampling strategy for extracting network information.

Lines 1--6 in Algorithm 2 show the procedure of node sequence sampling. As Fig.\ref{framework} shows we can obtain a series of node sequences using random walks.
For example, if we want a random walk of length $l = 5$ rooted at $A$ on the toy network, we may get the result of $\mathcal{W}=\{A,D,F,E,F\}$.
The other sequences are obtained similarly.

\begin{algorithm}[!ht]
\caption{Parameter learning}
\label{alg3}
\begin{algorithmic}[1]
\REQUIRE node representation $f$, pairs of nodes $E_p + E_n$,\\ train test split ratio $r'$
\ENSURE Evaluation results $val$
\STATE $D_t$, $D_p$ = DataSplit($E_p+E_n, f, r'$)
\STATE $p^{opt}=\text{SGD}(D_t)$

\FOR{$(u_j,v_j,y_j)\in D_p$}
	\STATE $tmp=(\mathbf{u}_j^T\mathbf{v}_j+p^{opt})/(|\mathbf{u}_j||\mathbf{v}_j|)$
	\STATE $\hat {y_j}=1/(1+e^{-tmp}$)
\ENDFOR
\STATE val = GetEvaluation($\mathbf{\hat y,y}$)
\RETURN val
\end{algorithmic}
\end{algorithm}

\subsubsection{Language model}
In order to get the representations of networks, the objective is to solving 
\begin{equation}
\max_f \sum_{v_i \in V}log~Pr(v_i|\mathbf{x}_{v_i}),
\end{equation}
where $\mathbf{x}_{v_i}=\frac{1}{2\lambda}\sum_{-\lambda \leqslant j \leqslant \lambda,j\neq 0} f(v_{i+j})$, $\lambda$ is the context size and $f$ is the mapping function from node to feature representations.
For $Pr(v_i|\mathbf{x}_{v_i})$, we can use softmax, which is a log-linear classification model, to get the posterior distribution of nodes.
However softmax involves the summation over all the node pairs and doing such computation for every training instance is very expensive, making it impractical to scale up to large networks.

\begin{table*}[!ht]
\renewcommand{\arraystretch}{1.3}
\centering
\caption{Basic topological information of the datasets. $|V|$ is the number of nodes in the network and $|E|$ is the total links. Avg. Degree denotes the average node degree. Avg.CC represents the average clustering coefficient which indicates the probability to be connected among neighbors of nodes. Diameter is the longest of all the calculated shortest paths in a network. Density is the ratio of $|E|$ to the number of possible edges.}
\label{nets}
%\resizebox{\textwidth}{!}{%
\begin{tabular}{ccrrrrrr}
\toprule
Type                    & Dataset      & $|V|$   & $|E|$    & Avg. Degree & Avg. CC & Diameter & Density                     \\ \midrule
\multirow{6}{*}{Dense} & C.elegans    & $297$   & $2,148$   & $14.47$       & $0.2924$  & $5$        & $4.89\times 10^{-2}$ \\
                        & PB           & $1,222$  & $16,714$  & $27.35$       & $0.3203$  & $8$        & $2.24\times10^{-2}$ \\
                        & Wiki-vote    & $7,066$  & $103,663$ & $28.50$       & $0.1419$  & $7$        & $4.15\times10^{-3}$ \\
                        & Email-enron  & $33,696$ & $183,831$ & $10.73$       & $0.4970$  & $11$       & $3.24\times10^{-4}$ \\
                        & Epinions     & $75,879$ & $508,837$ & $10.69$       & $0.1378$  & $15$       & $1.77\times10^{-4}$ \\
                        & Slashdot     & $77,360$ & $905,468$ & $14.13$       & $0.0555$  & $10$       & $3.03\times10^{-4}$ \\ \midrule
\multirow{4}{*}{Sparse} & Sexual       & $288$   & $291$    & $2.02$        & $0.0000$  & $37$       & $7.04\times10^{-3}$ \\
                        & Roadnet   & $2,092$  & $2,310$   & $2.21$        & $0.0065$  & $195$      & $1.06\times10^{-3}$ \\
                        & Power   & $4,941$  & $6,594$  &  $2.67$  &  $0.0801$  &  $46$     & $5.40\times10^{-4}$ \\
                        & Router       & $5,022$  & $6,258$   & $2.49$        & $0.0116$  & $15$       & $4.96\times10^{-4}$ \\
                        & p2p-Gnutella & $10,876$ & $39,994$  & $7.35$        & $0.0062$  & $10$       & $6.76\times10^{-4}$ \\ \bottomrule
\end{tabular}%
%}
\end{table*}

To solve this problem, an intuition is to limit the number of output vectors updated per training instance.
Thus hierarchical softmax\cite{2013arXiv1301.3781M} is proposed to improve the learning efficiency, which we adopt in this work.
In the end, we use stochastic gradient descent (SGD) techniques to optimize the objective function (lines 7--12 in Algorithm 2) to get the social representations of each node, i.e., $f(v_i)$, in the graph.
As illustrated in Fig.\ref{framework}, for each node in the toy network, we can get a $d$-dimensional representations associate with it.

\subsection{Similarity function learning}
For node pair $(u_i, v_i) \in D_S$, we use $\mathbf{u}_i$ and $\mathbf{v}_i$ as their features obtained from network representation learning.
Considering the distribution bias of real links among different geodesic distances, we propose a novel similarity function which is defined as
\begin{equation}
\label{kcos}
K(u_i,v_i)=\text{sim}(u_i,v_i)=\frac{\mathbf{u}_i^T \mathbf{v}_i + p}{\Vert \mathbf{u}_i \Vert \Vert \mathbf{v}_i \Vert}.
\end{equation}
We denote $a_i = \mathbf{u}_i^T \mathbf{v}_i$ and $b_i = \Vert \mathbf{u}_i \Vert \Vert \mathbf{v}_i \Vert$ for simplicity.
Then (\ref{kcos}) can be rewritten as 
\begin{equation}
\label{ksim}
K(u_i,v_i)=\frac{a_i+p}{b_i}.
\end{equation}
A logistic function is applied to mapping the node pair similarity to a value in $(0, 1)$, which is a probability indicating it belongs to the positive class or negative class.
We use $\hat{y_i}$ to denote this probability which is represented as
\begin{equation}
\label{logistic}
\hat{y_i}=\text{Logistic}(K)=\frac{1}{1+e^{-\frac{a_i+p}{b_i}}}.
\end{equation}
In order to measure the closeness between the predicted value and the true label, we select cross-entropy loss as our objective which is defined as
\begin{equation}
\label{obj}
C = -\frac{1}{N}\sum_{i=1}^N[y_i \text{log} \hat {y_i} + (1-y_i)\text{log} (1-\hat {y_i})],
\end{equation}
The stochastic gradient descent technique is used to get the optimal value of $p$ by minimizing $C$, it's updating rule can be written as 
\begin{equation}
p := p - \alpha \frac{dC}{dp},
\end{equation}
where
\begin{equation}
\frac{dC}{dp} = \frac{dC}{d\hat{y_i}}\frac{d\hat{y_i}}{dp}=\frac{1}{N}\sum_{i=1}^{N}(y_i-\hat{y_i})\frac{1}{b_i}.
\end{equation}

Algorithm \ref{alg3} shows the core part of the parameter learning process.
The training data set and test data set are first obtained through line 1 in Algorithm \ref{alg3}.
Then the optimal value of $p^{opt}$ is learned using SGD on the training data set $D_t$ (line 2).
The $p^{opt}$ is used to measure the similarity of node pairs in $D_p$ and we can get their probability of being connected through lines 3 to 6 in Algorithm \ref{alg3}.
Finally, the evaluation results are obtained through line 7.

\section{Experiments}\label{experiments}
%先介绍一下用到的数据，为了表明算法的通用性，我们在尽可能多、尽可能不一样的数据集上进行了测试。随后介绍了一些基准算法，包括传统的基于相似度的、最新的node2vec等
In this section, we first give a brief description of the datasets used in the experiment.
Next, we introduce the baseline models and evaluation metrics for link prediction.
Then, the experimental results are presented with a detailed analysis.
As the \textit{AdaSim} framework involves several parameters, lastly, we show how the different choices of these parameters affect the performance of link prediction.
\subsection{Datasets}\label{data}
So as to comprehensively evaluate the performance of our proposed link prediction algorithm, we use ten real-world datasets to conduct our experiments, and these datasets are commonly used in the link prediction domain.
These datasets come from various fields and their details are described as follows. 
\begin{itemize}
\item \textit{C.elegans}\cite{watts1998collective} is the neural network of the Caenorhabditis elegans worm. 
The nodes represent the neurons and the edges denote synapse or gap junction.

\item \textit{PB}\cite{Adamic:2005:PBU:1134271.1134277} is a network of hyperlinks between weblogs on United States politics.

\item \textit{Wiki-vote}\cite{snapnets} is a social network that contains all the Wikipedia voting data from the inception of Wikipedia till January 2008.
Nodes in the network represent wikipedia users and a directed edge from node $i$ to node $j$ represents that user $i$ voted on user $j$.

\item \textit{Email-enron}\cite{snapnets} is a communication network that covers all the email communication around half a million emails.
Nodes of the network are email addresses and if there is at least one email from address $i$ to address $j$, then they have a link between them.

\item \textit{Epinions}\cite{snapnets} is who-trust-whom online social network.
Members of the site of Epinions can decide whether to trust each other.
If user $i$ trust user $j$, then there is a link between them.

\item \textit{Slashdot}\cite{snapnets} is technology-related news website.
The network contains friend/foe links between the users of Slashdot.

\item \textit{Sexual}\cite{Bearman2004} is a well-known sexual contact network. This network is very sparse and has almost no closed triangles.

\item \textit{Roadnet}\cite{snapnets} denotes a road network of California, which is a typical sparse and treelike network.

\item \textit{Power} \cite{watts1998collective} is a traditional sparse network, which denotes the power grid of the western United States.

\item \textit{Router} is an Internet network of router-level collected by \textit{Rocketfuel Project}\cite{Spring:2004:MIT:973492.973494}.
\item \textit{p2p-Gnutella}\cite{snapnets} is a peer-to-peer file sharing network of Gnutella.
Nodes in the network represent hosts and edges represent connections among those hosts of Gnutella.
\end{itemize}

The basic topological information of these networks is listed in Table \ref{nets}, including the number of nodes and edges, average degree, average clustering coefficient, diameter and density of the network. We roughly divide the networks as dense and sparse based on the average degree and average clustering coefficient.
To sum up, we conduct experiments on networks with various properties, i.e., sparse and dense, small and large.
Thus the datasets can comprehensively reflect the characteristics of the proposed method \footnote{Different from other networks, $E^p$ has $100\%$ of the observed links for the highly sparse sexual contact network due to the small number of nodes.} .
\begin{table*}[]
\renewcommand{\arraystretch}{1.3}
\centering
\caption{AUC results of different algorithms. *For our algorithm, $E^p$ has $100\%$ of the observed links for the highly sparse sexual contact network, which moreover has only a small number of nodes.}
\label{auc-aupr}
\begin{tabular}{c|c|ccccccccc}
\toprule
        Type             & Dataset     & RA     & SI     & PA     & CCLP   & CN & HEI     & DeepWalk & Node2vec & AdaSim\\ \midrule
\multirow{6}{*}{Dense} & C.elegans    & 0.7229 & 0.6737 & 0.7214 & 0.7176 & 0.7078 & 0.6171 & 0.7622   & 0.7700   & \textbf{0.7758} \\
& PB           & 0.8734 & 0.8192 & 0.9020 & 0.8752 & 0.8710 & 0.7158 & 0.8765    & 0.8792   & \textbf{0.9243} \\
& Wiki-Vote    & 0.8843 & 0.8694 & 0.9569 & 0.8855 & 0.8842 & 0.7999 & 0.8515   & 0.8675   & \textbf{0.9657} \\
& Email-enron  & 0.8956 & 0.8894 & 0.9195 & 0.8912 & 0.8944 & 0.8440 & 0.9300   & 0.9301   & \textbf{0.9514} \\
& Epinions     & 0.8131 & 0.8114 & 0.9715 & 0.8092 & 0.8129 & 0.9089 & 0.8099   & 0.8177   & \textbf{0.9746} \\
&Slashdot 		 & 0.7012 & 0.7005 & 0.8564 & 0.6832 & 0.7009  & 0.7512 & 0.7241   & 0.7327   & \textbf{0.8756} \\\midrule
\multirow{4}{*}{Sparse}    & Sexual    & 0.4875 & 0.4875 & 0.4469 & 0.7481 & 0.5000 & 0.5031 & 0.7375	   & 0.8750   & \textbf{0.8750*}  \\
                     & Roadnet       & 0.5171 & 0.5171 & 0.3746 & 0.5000 & 0.5171 & 0.5325 & 0.7832   & 0.7909   & \textbf{0.8449} \\
                     & Power & 0.6177 &0.6177 &0.4925 &0.5000 &0.6177 &0.4963 &0.7533 & 0.7579 &{\textbf{0.7675}} \\
& Router       & 0.5557 & 0.5554 & 0.8919 & 0.5000 & 0.5556 & 0.8049 & 0.9074   & 0.9119   & \textbf{0.9315} \\
                    
& p2p-Gnutella & 0.5065 & 0.5065 & 0.7911 & 0.5018 & 0.5065 & 0.5689 & 0.5379   & 0.5572   & \textbf{0.7970} \\ \bottomrule
\end{tabular}
\end{table*}

\subsection{Baseline methods and evaluation metrics}
In order to validate the performance of our proposed algorithm, we compare \textit{AdaSim} against the following link prediction models.
\begin{itemize}
\item Common Neighbors (\textit{CN}). For node $u$, let $\Gamma (u)$ denote the set of neighbors of $u$. Two nodes, $u$ and $v$, have a high probability of being connected if they have many common neighbors\cite{Kossinets2006247, PhysRevE2001}.
The simplest way to measure this neighborhood overlap is by directly counting the number of common neighbors, i.e.,
\[
s_{uv}^{\text{CN}}=|\Gamma (u) \cap \Gamma(v)|.
\]
\item Resource Allocation (\textit{RA})\cite{Zhou2009}. For an unconnected node pair $u$ and $v$, it is assumed that $u$ can send some resources to $v$ by the medium of neighbors.
The similarity between $u$ and $v$ can be defined as the amount of resources received by $v$ from $u$, which described as
\[
s_{uv}^{\text{PA}}=\sum_{z\in \{\Gamma(u) \cap \Gamma(v)\}}\frac{1}{k_z},
\]
where $k_z$ is the degree of $z$.
\item Preferential Attachment (\textit{PA})\cite{Barabasi509}. Preferential attachment mechanism is used to generate random scale-free networks, in which the new links connecting to $u$ is proportional to $k_u$. Similarly, the probability that a new link connecting $u$ and $v$ is proportional to $k_u\times k_v$. The \textit{PA} similarity index is defined as
\[
s_{uv}^{\text{PA}}=k_u \times k_v.
\]
\item Salton Index (\textit{SI})\cite{Lu2011}. The other name of \textit{SI} is cosine similarity and is defined as
\[
s_{uv}^{\text{SI}}=\frac{|\Gamma(u) \cap \Gamma(v)|}{\sqrt{k_u \times k_v}}.
\]
\item Clustering Coefficient for Link Prediction (\textit{CCLP})\cite{Wu20161}. It is a similarity index with more local structural information considered. In this method, the local link information is conveyed by clustering coefficient of common neighbors.
\[
s_{uv}^{\text{CCLP}}=\sum_{z\in \{\Gamma(u) \cap \Gamma(v)\}}\frac{t_z}{k_z(k_z-1)/2},
\]
where $t_z$ is the number of triangles passing through node $z$.

\item Heterogeneity Index \cite{Shang2019}. This method is based on the network heterogeneity and the state-of-the-art for sparse and treelike networks. 
\[
s_{uv}^{\text{HEI}} = |k_u - k_v|^ \alpha,
\]
where $\alpha$ is a free heterogeneity exponent.

\item \textit{Node2vec}\cite{Grover2016}. This is a supervised way of link prediction using logistic regression. The features used in this method are generated through heuristic binary operators of node pair features which are learned from network embedding. There are two parameters, $p$ and $q$, to control the node sequences sampling. Note that when $p=q=1$, node2vec equals to DeepWalk\cite{Perozzi-2014}.
\end{itemize}

Beside \textit{Node2vec}, there are other approaches for unsupervised feature learning for graphs, such as spectral clustering\cite{Tang:2011:LSM:2036012.2036038} and LINE\cite{Tang:2015:LLI:2736277}.
We exclude them in this work since they have already been shown to be inferior to \textit{Node2vec}\cite{Grover2016}.
We also exclude other supervised methods, such as ensemble learning\cite{Lichtenwalter2010} and support vector machines\cite{al2006}. 
These methods can get relatively better performance but at the cost of high complexity, which is not our original attention.

We adopt the area under the receiver operating characteristic (AUC) to quantitatively evaluate the performance of link prediction algorithms.
The AUC value quantifies the probability that a randomly chosen missing link is given a higher score than a randomly chosen node pair without a link.
A higher score means better performance.

%***********************************Figure BEGIN***************************************
\begin{figure*}[]
\centering
\renewcommand{\arraystretch}{1.3}
\subfloat[][C.elegans]{\includegraphics[width=0.3\textwidth]{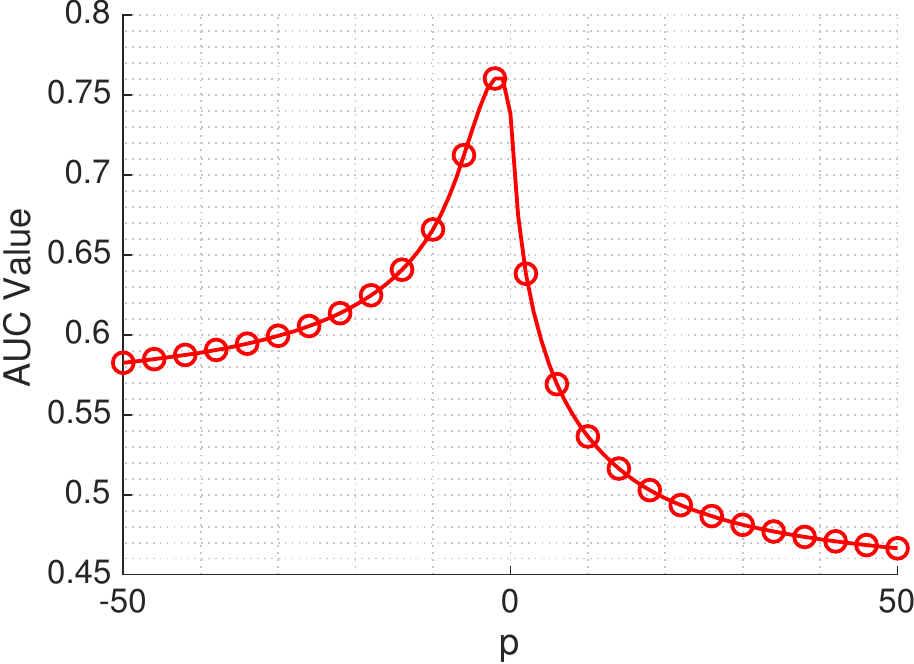}\label{celegans}}\hfill
\subfloat[][Router]{\includegraphics[width=0.3\textwidth]{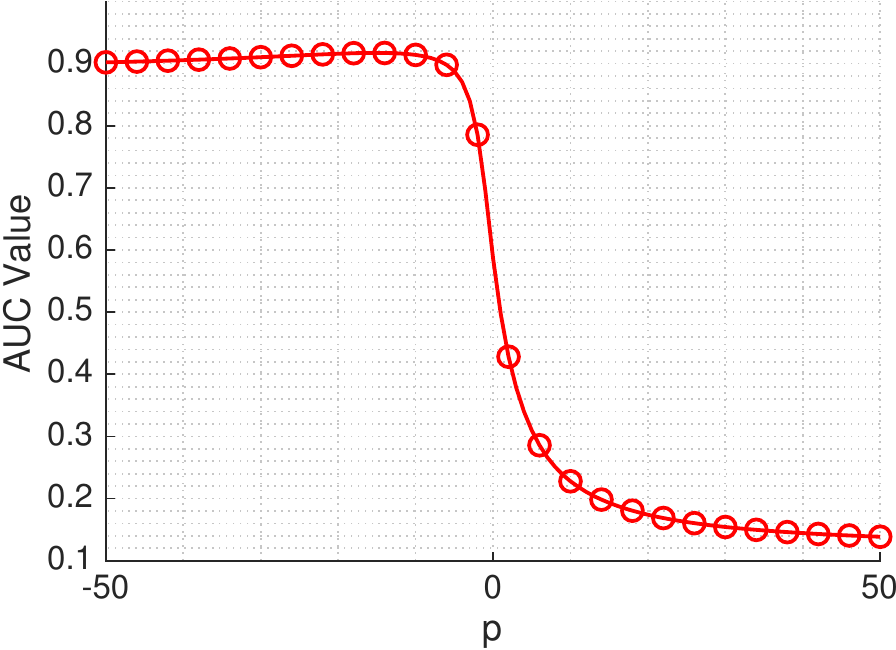}\label{router}}\hfill
\subfloat[][Wiki-vote]{\includegraphics[width=0.3\textwidth]{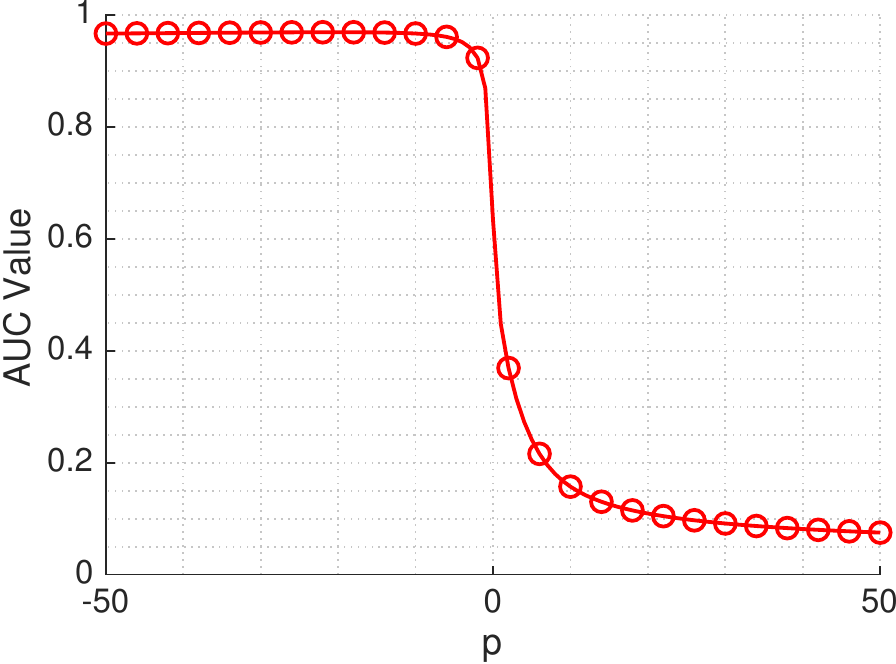}\label{wiki}}
\caption{The change of AUC with different choices of $p$.}
\label{p_performance}
\end{figure*}
%***********************************Figure END*****************************************

\subsection{Experimental results}

In order to obtain the following results, we set the parameters in line with the typical values in \cite{Grover2016}. 
That is, $d=128, k=10, l = 80, \lambda = 10$, and the optimization is run for a single epoch.
Fifty percent of the edges are removed and treated as positive examples.
The negative node pairs which have no edge connecting them are randomly sampled from the network.
For the two parameters, $p$ and $q$, in \textit{Node2vec}, they are selected through a grid search over $p,q \in \{0.25,0.5,1,2\}$.
After the dataset is prepared we use ten-fold cross validation to evaluate the performance.
For the sake of objectivity, the experiment is repeated ten times on each dataset and the average results are reported in Table \ref{auc-aupr}.

A general observation we can draw from these results is that the proposed link prediction algorithm, \textit{AdaSim}, can obtain better performance than all the baseline methods on all datasets.
More specifically, the unsupervised similarity-based link prediction methods achieve relatively lower value than those supervised ones, since the label information is not leveraged to boost model performance.
But the \textit{PA} predictor achieves competitively results compared with \textit{AdaSim} and even better than \textit{Node2vec} on five out of eleven datasets.
This is because preferential attachment is one of the key features in generating power law scale-free networks.
It reflects the mechanism of network evolution that involves the addition of new nodes and edges.
Thus it can obtain better performance on link prediction problems. But similarity-based based link prediction methods perform extremely worse when the network is sparse since limited or no closed triangular structure exists in these networks.

Among all the supervised link prediction methods, \textit{AdaSim} outperforms both \textit{DeepWalk} and \textit{Node2vec} in all the eleven networks with gain ratios of different scales.
The gain ratio varies from 0.75\% to 43.04\% in the AUC values compared with \textit{Node2vec}.

%Though the obtained improvements varies in evaluation metrics, it can demonstrate the effectiveness of \textit{AdaSim} for link prediction problem.
%In other words, by defining a flexible similarity function with a penalty $p$ to balance link generation probability among different geodesic distances, the various link generation patterns can be captured by \textit{AdaSim} and which can furthermore enhance the link prediction performances.

For intuitively show the influence of penalty $p$ on link prediction performance, $p$ is set to specific values from $-val$ to $val$ with fixed increment $a$ (here $val=50, a=1$ for demonstration) and display the results of AUC on three datasets, i.e., \textit{C.elegans}, \textit{Router} and \textit{Wiki-vote}, in Fig.\ref{p_performance}.
Notice that $p=0$ corresponds to the original cosine similarity measurement.
It can be clearly seen, from Fig.\ref{p_performance}, that the results of AUC are considerably affected by the value of $p$.
Compared with the rigid cosine similarity, our proposed \textit{AdaSim} can substantially improve the link prediction performance.
This also verifies our empirical findings in section \ref{evidence} that different networks have different link formation patterns, thus a flexible and adaptive similarity function for link prediction is needed to capture these various patterns.

% \usepackage{graphicx}
\begin{table}[]
\centering
\renewcommand{\arraystretch}{1.3}
\caption{Running time (s) comparisons among different algorithms on three representative sparse networks.}
\label{tab:wall}
\resizebox{0.45\textwidth}{!}{%
\begin{tabular}{llllll}  \toprule
{Dataset}      & {CN} & {CCLP}   & {HEI}    & {Node2vec} & {AdaSim} \\ \midrule
{Sex}          & {0.0020}                 & {0.0015} & {0.0010} & {0.0240}   & {0.0045} \\
{Power}        & {0.2280}                 & {0.0475} & {0.0255} & {0.2735}   & {1.2755} \\
{p2p-Gnutella} & {4.4310}                 & {0.8085} & {0.3205} & {1.6615}   & {4.0910} \\ \bottomrule   
\end{tabular}%
}
\end{table}

%***********************************Figure BEGIN***************************************
\begin{figure}[]
\centering
\includegraphics[width=0.4\textwidth]{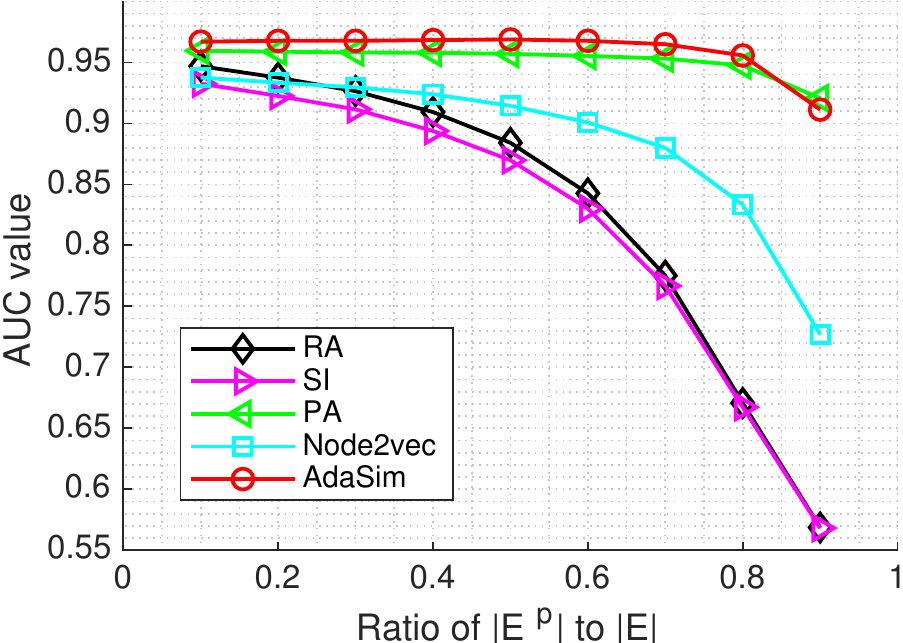}\label{auc-ratio}
\caption{The performance of various link prediction methods on networks with different sparsity.}
\label{sparsity_test}
\end{figure}
%***********************************Figure END*****************************************

We select three representative sparse networks, that is, Sex (small), Power (medium), and p2p-Gnutella (large), and report the wall-clock time of CN, CCLP, HEI, Node2vec, and AdaSim in Table \ref{tab:wall}. We can observe from this table that with the increase of network size, the prediction time needed for all algorithms also increases. Besides, the learning-based algorithms usually take more time than similarity-based ones since vector-vector multiplication takes more time than simply calculating the neighborhood information of two nodes. 
Though our algorithm requires more time to predict, it has also achieved considerable performance gains, as we explained above.

%***********************************Figure BEGIN***************************************
\begin{figure*}[]
\centering
\subfloat[][]{\includegraphics[width=0.33\textwidth]{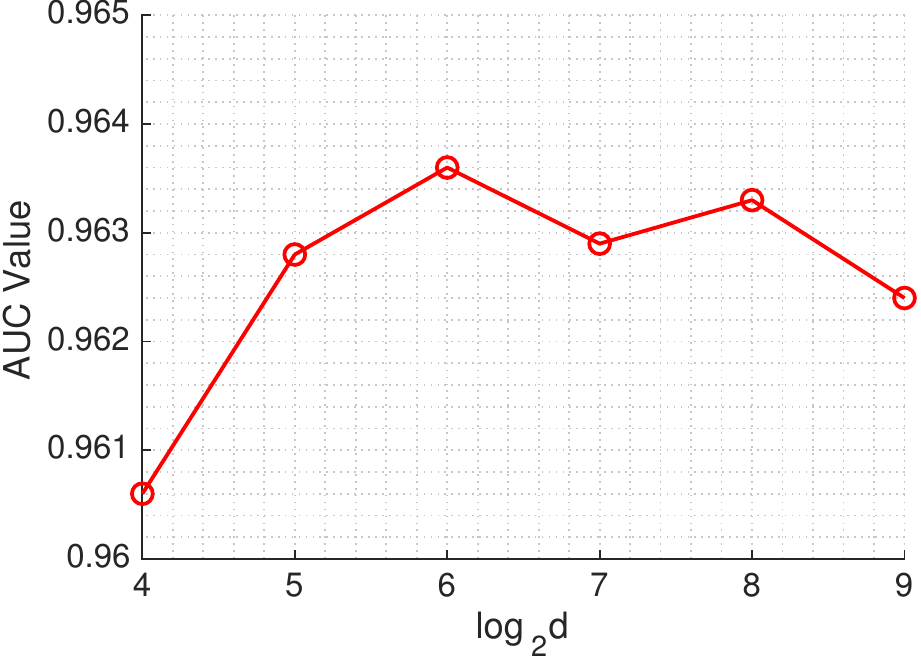}\label{d_size}}\hfill
\subfloat[][]{\includegraphics[width=0.33\textwidth]{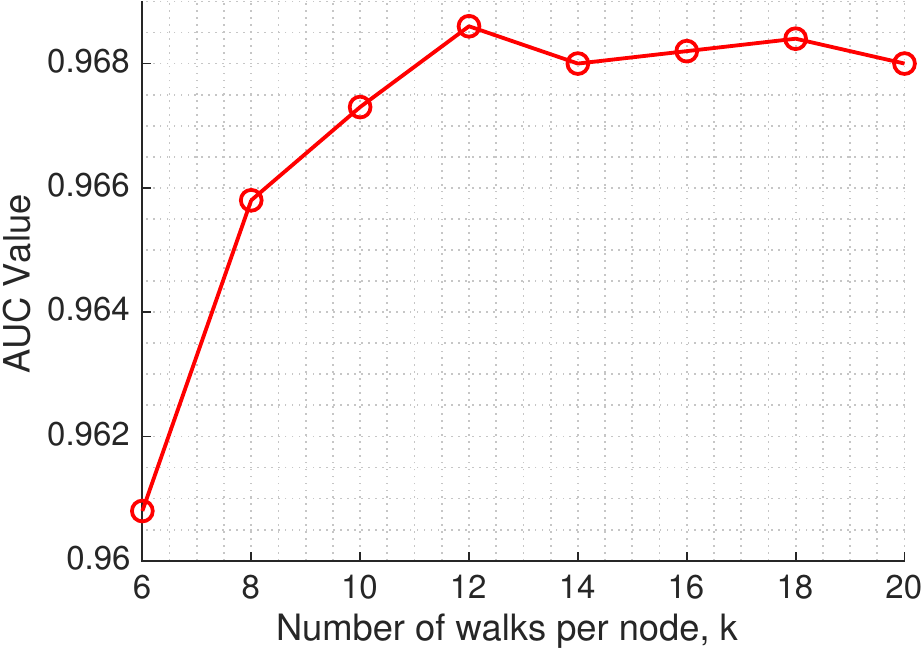}\label{l_size}}\hfill
\subfloat[][]{\includegraphics[width=0.33\textwidth]{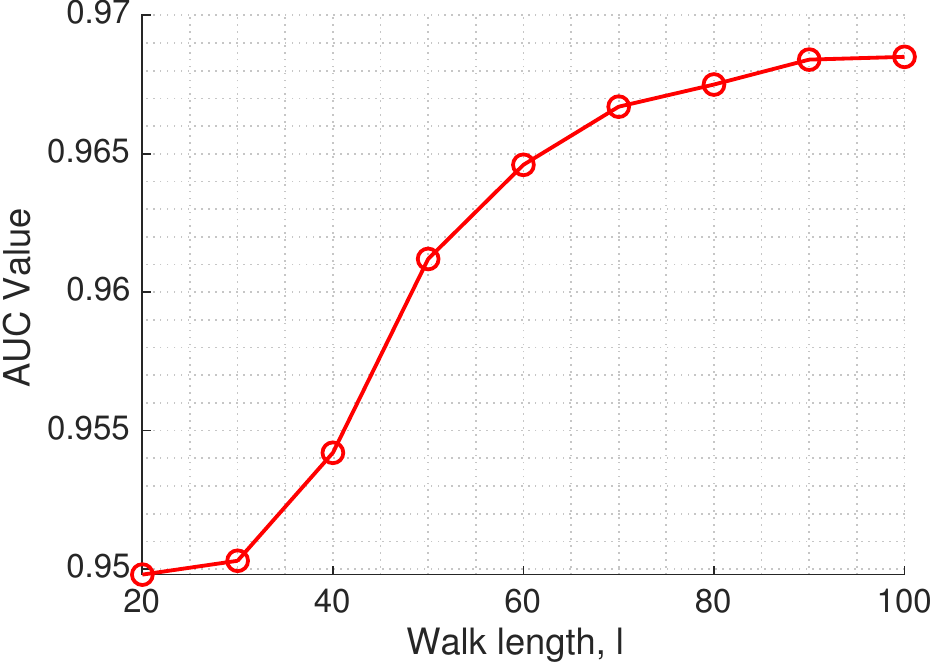}\label{k_size}}
\caption{Parameter sensitivity of \textit{AdaSim}}
\label{parameters}
\end{figure*}
%***********************************Figure END*****************************************

\subsection{Performance on networks with different sparsity}\label{xx}
Networks in the real world are often sparse, we only know very limited information about the interactions among the nodes.
For example, 80\% of the molecular interactions in cells of yeast and 99.7\% of human are still unknown\cite{Lu2011}.
A good link prediction method should have robust performance on networks with different sparsity.

We change the sparsity of the networks by randomly removing a certain percent of links in the original network, then follow the aforementioned experiment setup to report the results of different methods.
The results on the Wiki-Vote dataset are displayed in Fig.\ref{sparsity_test}.
Only four baseline methods are listed in the figure since \textit{CN,CCLP} and \textit{DeepWalk} performs similarly with \textit{RA} and \textit{Node2vec}, respectively.

It can be seen from the results that  the AUC values decrease with the increase of removed edge ratio since it is becoming more and more challenging to characterize node similarity using the information of network topology.
The similarity-based methods perform well when the removed edge ratio is relatively small.
But they degrade very quickly and give their way to \textit{Node2vec} and \textit{AdaSim} except the \textit{PA} predictor, which have competitively performance but still insuperior to \textit{AdaSim}.
\textit{AdaSim} performs consistently well and robust to different sparsity conditions of networks.
Even when eighty percent of the edges are removed, the \textit{AdaSim} can still hold the performance around 0.95 in terms of AUC.
Overall, \textit{AdaSim} is not only robust to different network conditions but also achieves better performance than baselines.

\subsection{Parameter sensitivity}

There are several parameters involved in the \textit{AdaSim} algorithm and in Fig.\ref{parameters}, we examine how the different choices of parameters influence the performance of \textit{AdaSim} on the Wiki-Vote dataset.
Except for the parameter being tested, all other parameters assume default values.

We measure the AUC as a function of the representation dimension $d$, walk length $l$ and the number of walks per node $k$.
We observe that the dimension of learning representations for nodes has limited effects on link prediction performance.
With the increase of dimensionality, the AUC values increase slightly and turn to saturate when $d$ reaches to 128.
It can also be observed that a larger $l$ and $k$ will improve the performance, this is because more neighborhood information of the seed node is included in the representation learning process, the node similarities can be captured more precisely.

\section{Conclusion}\label{conclusion}
In this work, we focus on the link prediction problem with features obtained from network embedding.
As the edge features generated through heuristic binary operators are an information-loss projection of the original node features, we have quantitatively given the evidence of inconsistency between heuristic edge features and learned ones.
Moreover, we have developed a novel link prediction framework \textit{AdaSim} by introducing an adaptive similarity function to deal with the inflexible of cosine similarity, especially for sparse or treelike networks.
\textit{AdaSim} first learns node representations of networks by solving a graph-based objective function, then adds a penalty parameter, $p$, on the original similarity function.
At last, the optimal value of $p$ is learned through supervised learning.
The proposed \textit{AdaSim} is flexible thus is adaptive to data distribution and can capture the various link formation mechanisms of different networks.
We conducted experiments using publicly available real-world network datasets, and extensively compared \textit{AdaSim} with seven well-established representative baseline methods.
The results show that \textit{AdaSim} achieves better performance than state-of-the-art algorithms on all datasets.
It is also robust to the sparsity of the networks and obtains competitive performance with even a large fraction of edges are missing.

\section{Acknowledgements}
This work is supported by National Natural Science Foundation of China (61901247, 61803047), Natural Science Foundation of Shandong Province ZR2019BF032, Major Project of The National Social Science Foundation of China (19ZDA149, 19ZDA324) and Fundamental Research Funds for the Central Universities (14370119, 14390110).
\section{Data Availability}
All datasets can be obtained from the corresponding author upon request.

\section{Conflicts of Interest}
The authors declare no competing interests.

%% use section* for acknowledgment
%\ifCLASSOPTIONcompsoc
%  % The Computer Society usually uses the plural form
%\else
%  % regular IEEE prefers the singular form
%\fi

\ifCLASSOPTIONcaptionsoff
  \newpage
\fi

\bibliographystyle{IEEEtran}
\bibliography{IEEEabrv,adasim}

%\begin{IEEEbiography}{Chuanting Zhang}
%\end{IEEEbiography}

\end{document}